\documentclass[11pt,journal,draftcls,onecolumn,peerreviewca]{IEEEtran}

\usepackage{amsfonts}
\usepackage{amsmath}
\usepackage{amssymb}

\usepackage{algorithmic}
\usepackage{algorithm}
\usepackage{graphicx}
\usepackage{color, soul}
\usepackage{stfloats}
\usepackage{setspace}
\usepackage{algorithm}
\usepackage{algorithmic}
\usepackage{float}
\usepackage{verbatim}
\usepackage{cite}
\usepackage{url}
\usepackage{bm}
\usepackage{stmaryrd}
\usepackage{multicol}
\usepackage{stfloats}
\usepackage[amsmath,thmmarks]{ntheorem}
\usepackage{theorem}
\usepackage[justification=centering]{caption}

\theoremheaderfont{\sc}\theorembodyfont{\upshape}
\theoremstyle{nonumberplain}
\theoremseparator{}
\theoremsymbol{\rule{1ex}{1ex}}

\newcommand{\red}{\textcolor[rgb]{1.00,0.00,0.00}}

\begin{document}

\title{Bilinear Adaptive Generalized Vector Approximate Message Passing}
\author{Xiangming Meng and Jiang Zhu  \thanks{Xiangming Meng  is with Huawei Technologies, Co. Ltd., Shanghai, China. (e-mail: mengxm11@gmail.com). Jiang Zhu is with the key laboratory of ocean observation-imaging testbed of Zhejiang Province, Ocean College, Zhejiang University, No.1 Zheda Road, Zhoushan, China, 316021 (e-mail:jiangzhu16@zju.edu.cn). The corresponding author of this work is Jiang Zhu (e-mail: jiangzhu16@zju.edu.cn).
}}

\date{}
\maketitle
\begin{abstract}
This paper considers the generalized bilinear recovery problem which aims to jointly recover the vector $\mathbf b$ and the matrix $\mathbf X$ from componentwise nonlinear measurements ${\mathbf Y}\sim p({\mathbf Y}|{\mathbf Z})=\prod\limits_{i,j}p(Y_{ij}|Z_{ij})$, where ${\mathbf Z}={\mathbf A}({\mathbf b}){\mathbf X}$, ${\mathbf A}(\cdot)$ is a known affine linear function of $\mathbf b$, and $p(Y_{ij}|Z_{ij})$ is a scalar conditional distribution which models the general output transform. A wide range of real-world applications, e.g., quantized compressed sensing with matrix uncertainty, blind self-calibration and dictionary learning from nonlinear measurements, one-bit matrix completion, joint channel and data decoding, etc., can be cast as the generalized bilinear recovery problem. To address this problem, we propose a novel algorithm called the Bilinear Adaptive Generalized Vector Approximate Message Passing (BAd-GVAMP),  which extends the recently proposed Bilinear Adaptive Vector AMP (BAd-VAMP) algorithm to incorporate arbitrary distributions on the output transform. Numerical results on various applications demonstrate the effectiveness of the proposed BAd-GVAMP algorithm.
\end{abstract}
{\bf{keywords}}:
Generalized bilinear model, approximate message passing, expectation propagation,
expectation maximization, dictionary learning, self-calibration, matrix factorization.
\section{Introduction}
In this work, we consider the generalized bilinear recovery problem: jointly estimate the vector $\mathbf b$ and the matrix $\mathbf X$ from componentwise and probabilistic  measurements ${\mathbf Y}\sim p({\mathbf Y}|{\mathbf Z})=\prod\limits_{i,j}p(Y_{ij}|Z_{ij})$, where ${\mathbf Z}={\mathbf A}({\mathbf b}){\mathbf X}$, ${\mathbf A}(\cdot)$ is a known affine linear function of $\mathbf b$ (i.e., ${\mathbf A}({\mathbf b})={\mathbf A}_0+\sum\limits_{i=1}^Gb_i{\mathbf A}_i$ with known matrices ${\mathbf A}_i$.). This problem arises in a wide range of applications in the field of signal processing and computer science. For example, compressed sensing under matrix uncertainty \cite{SpaCog, Rosenbaum, ParkerMU, KrzakalaMU},  matrix completion \cite{MC, Tanaka, Zdeborov}, robust principle component analysis (RPCA) \cite{RPCA}, dictionary learning \cite{Parker1, Parker2}, joint channel and data decoding \cite{JCD, JCDWu, JCDWen} can all be formulated as generalized bilinear recovery problem. Generally the scalar conditional distribution $p(Y_{ij}|Z_{ij})$ models arbitrary componentwise measurement process in a probabilistic manner. Specially,  $p(Y_{ij}|Z_{ij})={\mathcal N}(Y_{ij};Z_{ij},\gamma_w^{-1})$ corresponds to the scenario of linear measurements, i.e., ${\mathbf Y}={\mathbf A}({\mathbf b}){\mathbf X}+{\mathbf W}$, where ${\mathcal N}(x;a,\gamma^{-1})$ denotes a Gaussian distribution with mean and variance being $a$ and $\gamma^{-1}$. In practice, however, the measurements are often obtained in a nonlinear way. For example, quantization is a common nonlinear measurement process in analog-to-digital converter (ADC) that maps the input signal from continuous space to discrete space, which has been widely used in (one-bit) compressed sensing \cite{BIHT}, millimeter massive multiple input multiple output (MIMO) system\cite{massMIMO},\cite{Mo}, etc. As a result, it is  of high significance to  study the generalized bilinear recovery problem.

There has been extensive research on this active field in the past few years, including the convex relaxation  methods \cite{SC, Convex}, variational methods \cite{SBLMC}, approximate message passing (AMP) methods such as bilinear generalized AMP (BiGAMP) \cite{Parker1, Parker2} and parametric BiGAMP (PBiGAMP) \cite{PbiGAMP}, etc. It was shown that the AMP based methods are competitive in terms of phase transition and computation time \cite{Parker1, Parker2, PbiGAMP, KrzakalaMC}. However, as the measurement matrix deviates from the i.i.d. Gaussian, the AMP may diverge \cite{Bayati1, Bayati2}. To improve convergence of AMP, vector approximate message passing (VAMP)\cite{VAMP} and orthogonal AMP (OAMP) \cite{OAMP} have been recently proposed, which achieve good convergence performance for any right-rotationally invariant measurement matrices and can be rigorously characterized by the scalar state evolution. For the generalized linear model, AMP is extended to generalized approximate message passing (GAMP) \cite{GAMPconf, GAMParxiv}. Later, generalized VAMP \cite{GVAMP} and generalized expectation consistent algorithm \cite{GrEC} are proposed to handle a class of right-rotationally invariant measurement matrices. In \cite{GSBL}, a unified Bayesian inference framework is provided and some insights into the relationship between AMP (VAMP) and GAMP (GVAMP) are presented. Due to the improved convergence of VAMP over AMP on general measurement matrices, many works have been done to extend VAMP to deal with the bilinear recovery problem \cite{LVAMP, RVAMP}. In \cite{LVAMP}, lifted VAMP is proposed for standard bilinear inference problem such as compressed sensing with matrix uncertainty and self-calibration. However, lift VAMP suffers from high computational complexity since the number of unknowns increases significantly, especially when the number of original variables is large. To overcome the computation issue, the bilinear adaptive VAMP (BAd-VAMP) has been  proposed very recently in \cite{BAdVAMP} which avoids lifting and instead builds on the adaptive AMP framework \cite{AdAMP, Fletcher2017}. Nevertheless, BAd-VAMP is only applicable to linear measurements which limits its usage in the generalized bilinear recovery problem.

In this paper, we propose a new algorithm called the bilinear adaptive generalized vector AMP (BAd-GVAMP),  which extends the BAd-VAMP \cite{BAdVAMP} from linear measurements to nonlinear measurements. Specifically, a novel factor graph representation of the generalized bilinear problem is first proposed by incorporating the Dirac delta function. Then, by using the expectation propagation (EP) \cite{Minka}, we decouple the original generalized bilinear recovery problem into two modules: one module performs componentwise minimum mean square error (MMSE) estimate while the other performs BAd-VAMP with some slight modification of the message passing schedule. Furthermore, the messages exchanging between the two modules are derived to obtain the final BAd-GVAMP. Interestingly, BAd-GVAMP reduces to the BAd-VAMP under linear measurements. Numerical results are conducted for quantized compressed sensing with matrix uncertainty, self-calibration as well as structured dictionary learning from quantized measurements, which demonstrates the effectiveness of the proposed algorithm.
{\subsection{Notation}}
Let ${\mathcal N}({\mathbf x};{\boldsymbol \mu},{\boldsymbol \Sigma})$ denote a Gaussian distribution of the random variable $\mathbf x$ with mean ${\boldsymbol \mu}$ and covariance matrix ${\boldsymbol \Sigma}$. Let $(\cdot)^{\text T}$, \red{${\Vert {\cdot} \Vert}_{\text F}$}, ${\Vert {\cdot} \Vert}$, $p(\cdot)$ and $\delta(\cdot)$ denote the transpose operator, the Frobenius norm, the $l_2$ norm, the probability density function (PDF) and the Dirac delta function, respectively. Let $<{\mathbf x}>$ denote the average $<{\mathbf x}>=\sum\limits_{i=1}^Nx_i/N$ for ${\mathbf x}\in {\mathbb R}^N$.
\section{Problem Setup}
Consider the generalized bilinear recovery problem as follows: jointly estimate the matrix ${\mathbf X}\in {\mathbb R}^{N\times L}$ and the  parameters ${\boldsymbol \Theta}\triangleq \{{\boldsymbol \theta }_X,{\boldsymbol \theta}_A,{\boldsymbol \theta}_Y\}$ from the componentwise probabilistic measurements ${\mathbf Y}\in{\mathbb R}^{M\times L}$, i.e.,
\begin{subequations}\label{genmodel}
\begin{align}
&{\mathbf X}\sim p({\mathbf X};{\boldsymbol \theta}_X)=\prod\limits_{i,j}p(x_{ij};{\boldsymbol \theta}_X)=\prod\limits_{l=1}^Lp({\mathbf x}_l;{\boldsymbol \theta}_X),\\
&{\mathbf Z}={\mathbf A}({\boldsymbol \theta}_A){\mathbf X},\\
&{\mathbf Y}\sim p({\mathbf Y}|{\mathbf Z};{\boldsymbol \theta}_Y)=\prod\limits_{i,j}p(Y_{ij}|Z_{ij};{\boldsymbol \theta}_Y)=\prod\limits_{l=1}^Lp({\mathbf y}_l|{\mathbf z}_l;{\boldsymbol \theta}_Y),
\end{align}
\end{subequations}
where ${\mathbf Y}$ denotes the nonlinear observations, ${\mathbf A}(\cdot)\in{\mathbb R}^{M\times N}$ is a known matrix-valued linear function parameterized by the unknown vector ${\boldsymbol \theta}_A$, $p({\mathbf X};{\boldsymbol \theta}_X)$ is the prior distribution of ${\mathbf X}$ parameterized by ${\boldsymbol \theta}_X$, $p({\mathbf Y}|{\mathbf Z};{\boldsymbol \theta}_Y)$ is the componentwise probabilistic output distribution conditioned on ${\mathbf Z}$ and parameterized by ${\boldsymbol \theta}_Y$. Given the above statistical model, the goal is to compute the maximum likelihood (ML) estimate of ${\boldsymbol \Theta}$ and the MMSE estimate of ${\mathbf X}$, i.e.,
\begin{align}\label{thetaML}
&\hat{\boldsymbol \Theta}_{\text {ML}}= \underset{\boldsymbol \Theta}{\operatorname{argmax}}~p_{\mathbf Y}({\mathbf Y};{\boldsymbol \Theta}),\\
&\hat{\mathbf X}_{\text {MMSE}}= {\text E}[{\mathbf X}|{\mathbf Y};\hat{\boldsymbol \Theta}_{\text {ML}}],
\end{align}
where $p_{\mathbf Y}({\mathbf Y};{\boldsymbol \Theta})=\int p({\mathbf X};{\boldsymbol \Theta})p({\mathbf Y}|{\mathbf X};{\boldsymbol \Theta}){\text d}{\mathbf X}$ is the likelihood function of ${\boldsymbol \Theta}$ and the expectation is taken with respect to the posterior probability density distribution
\begin{align}\label{postpdf0}
p({\mathbf X}|{\mathbf Y};\hat{\boldsymbol \Theta}_{\text {ML}})=\frac{p({\mathbf X},{\mathbf Y};\hat{\boldsymbol \Theta}_{\text {ML}})}{p({\mathbf Y};\hat{\boldsymbol \Theta}_{\text {ML}})},
\end{align}
where $p({\mathbf X},{\mathbf Y};\hat{\boldsymbol \Theta}_{\text {ML}})$ is
\begin{align}\label{jointpdf}
p({\mathbf X},{\mathbf Y};\hat{\boldsymbol \Theta}_{\text {ML}})=p({\mathbf X};\hat{\boldsymbol \Theta}_{\text {ML}})p({\mathbf Y}|{\mathbf X};\hat{\boldsymbol \Theta}_{\text {ML}}).
\end{align}
However, exact ML estimate of ${\boldsymbol \Theta}$ and exact MMSE estimate of ${\mathbf X}$ is intractable due to high-dimensional integration. As a result, approximate methods need to be designed in practice.

\section{Biliear Adaptive Generalized VAMP}
\begin{figure}
  \centering
  \includegraphics[width=8cm]{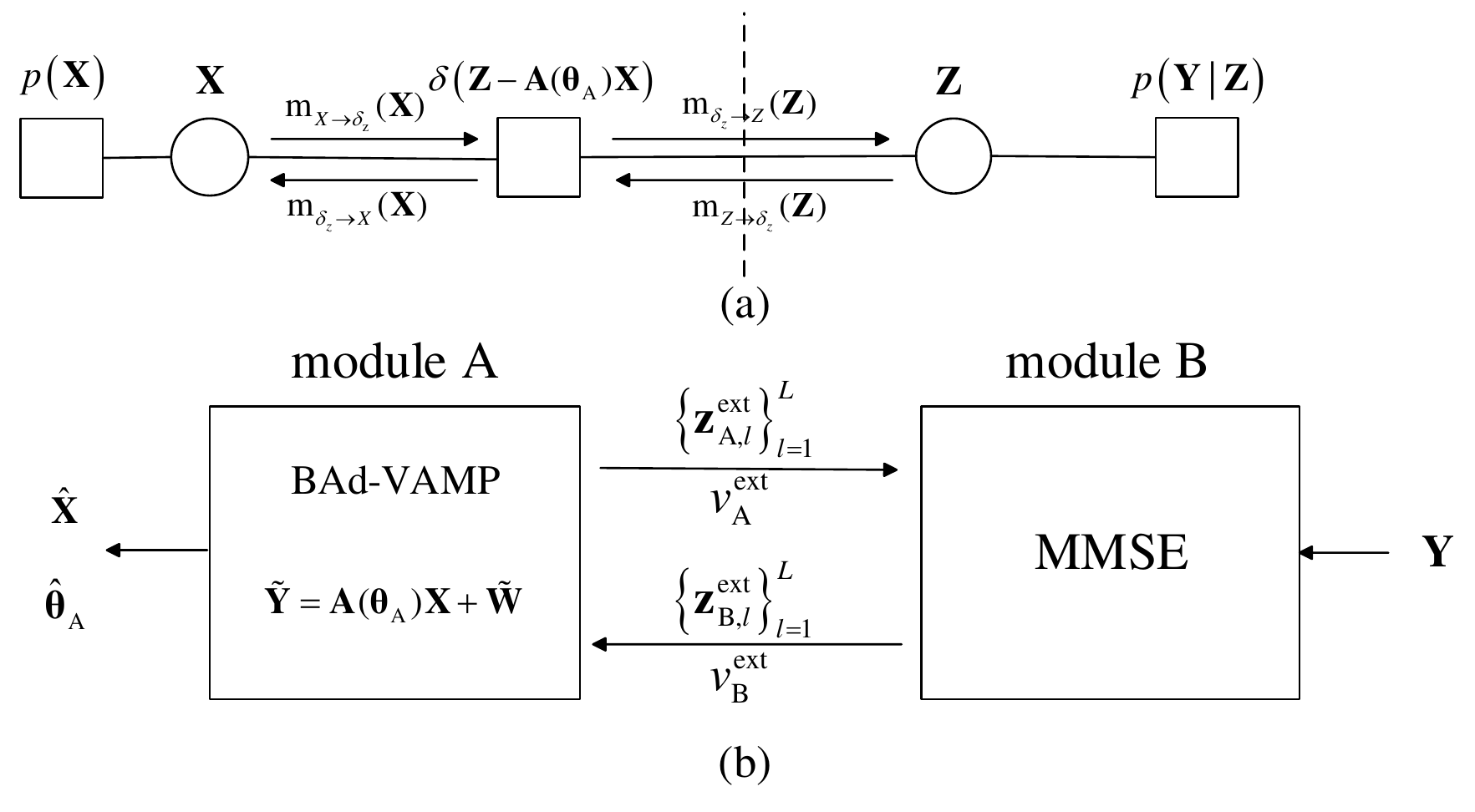}\\
  \caption{The factor graph and inference module of the BAd-GVAMP algorithm.}\label{module}
\end{figure}

In this section, we propose an efficient algorithm to approximate the ML estimate of ${\boldsymbol \Theta}$ and MMSE estimate of ${\mathbf X}$. The resultant BAd-GVAMP algorithm is an extension of BAd-VAMP from linear measurements to nonlinear measurements. To begin with, we first present  a novel factor graph representation of the statistical model. By introducing a hidden variable ${\mathbf Z}$ and a Dirac delta function $\delta(\cdot)$, the joint distribution in (\ref{jointpdf}) can be equivalently factored as
\begin{align}\label{jointpdfv2}
p({\mathbf X},{\mathbf Y};{\boldsymbol \Theta})= p({\mathbf X};{\boldsymbol \theta}_X)p({\mathbf Y}|{\mathbf Z};{\boldsymbol \theta}_Y)\delta({\mathbf Z}-{\mathbf A}({\boldsymbol \theta}_A){\mathbf X}).
\end{align}
The corresponding factor graph of (\ref{jointpdfv2}) is shown in Fig. \ref{module} (a). The circles and squares denote the variable and factor node, respectively. Such alternative factor graph representation plays a key role in the design of our approximate estimation algorithm. Now we will derive the BAd-GVAMP algorithm based on the presented factor graph  in  Fig. \ref{module} (a) and the EP \cite{Minka}. As one kind of approximate inference methods, EP approximates the target distribution p with an exponential family distribution (usually Gaussian) set ${\boldsymbol \Phi}$ which minimizes the Kullback-Leibler (KL) divergence ${\text {KL}}(p||q)$, i.e., $q = {\text {Proj}}(p) = \underset{q\in {\boldsymbol \Phi}}{\operatorname{argmin}}~ {\text {KL}}(p||q)$. For Gaussian distribution set ${\boldsymbol \Phi}$, EP amounts to moment matching, i.e., the first and second moments of distribution $q$ matches those of the target distribution.  For more details of EP and its relation to AMP methods, please refer to \cite{Minka, VAMP, EPmeng, EPzhang, WuEP, EC}.

To address the generalized bilinear recovery problem\red{, s}pecifically, we choose the projection set ${\boldsymbol \Phi}$ to be Gaussian with scalar covariance matrix, i.e., diagonal matrix whose diagonal elements are equal \footnote{Note that general diagonal matrix can also be used.}. Then, using EP on the factor graph  in  Fig. \ref{module}, we decouple the original generalized bilinear recovery problem  into two modules: the componentwise MMSE module and the BAd-VAMP module. The two modules interact with each other iteratively with extrinsic messages exchanging between them. The detailed derivation of BAd-GVAMP is presented as follows.
\subsection{Componentwise MMSE module}

Suppose that in the $t$-th iteration,  the message ${m}_{\delta_z\rightarrow {\mathbf Z}}^{t-1}({\mathbf Z})=\{m_{\delta_z\rightarrow {\mathbf z}_l}^{t-1}({\mathbf z}_l)\}_{l=1}^L$  from  factor node $\delta({\mathbf Z}-{\mathbf A}({\boldsymbol \theta}_A){\mathbf X})$ to  variable node ${\mathbf Z}$ follows Gaussian distribution, i.e.,
\begin{align}\label{init1}
m_{\delta_z\rightarrow {\mathbf z}_l}^{t-1}({\mathbf z}_l)={\mathcal N}({\mathbf z}_l;{\mathbf z}_{{\text A},l}^{\text {ext}}(t-1),{v}_{{\text A}}^{\text {ext}}(t-1){\mathbf I}_M),
\end{align}
where $\delta_z$ refers to the factor node $\delta({\mathbf Z}-{\mathbf A}({\boldsymbol \theta}_A){\mathbf X})$. According to EP, the message $m_{{\mathbf z}_l\rightarrow \delta_z}^t({\mathbf z}_l)$ from variable node ${\mathbf Z}$ to the factor node $\delta({\mathbf Z}-{\mathbf A}({\boldsymbol \theta}_A){\mathbf X})$ can be calculated as
\begin{align}
m_{{\mathbf z}_l\rightarrow \delta_z}^t({\mathbf z}_l)&\propto \frac{{\text {Proj}}[p({\mathbf y}_l|{\mathbf z}_l;{\boldsymbol \theta}_Y^{t-1})m_{\delta_z\rightarrow {\mathbf z}_l}^{t-1}({\mathbf z}_l)]}{m_{\delta_z\rightarrow {\mathbf z}_l}^{t-1}({\mathbf z}_l)}\label{delta0a}\\
&\triangleq \frac{{\text {Proj}}[q_{\text B}^t({\mathbf z}_l)]}{m_{\delta_z\rightarrow {\mathbf z}_l}^{t-1}({\mathbf z}_l)},\label{delta0}
\end{align}
where $\propto$ denotes identity up to a normalizing constant.
First, we perform componentwise MMSE and obtain the posterior means and variances of ${\mathbf z}_l$ as
\begin{align}
&{\mathbf z}_{{\text B},l}^{\text {post}}(t)={\text E}[{\mathbf z}_l|q_{\text B}^t({\mathbf z}_l)],\label{comb_means}\\
&{v}_{{\text B},l}^{\text {post}}(t)=<{\text {Var}}[{\mathbf z}_l|q_{\text B}^t({\mathbf z}_l)]>\label{comb_vars},
\end{align}
where ${\text E}[\cdot|q_{\text B}^t({\mathbf z}_l)]$ and ${\text {Var}}[\cdot|q_{\text B}^t({\mathbf z}_l)]$ are the mean and variance operations taken (componentwise) with respect to the distribution $\propto q_{\text B}^t({\mathbf z}_l)$ (\ref{delta0}). Then the posterior variances ${v}_{{\text B},l}^{\text {post}}(t)$ are averaged over $l$ which yields
\begin{align}\label{postzb}
{v}_{\text B}^{\text{post}}(t)=\sum\limits_{l=1}^L{v}_{{\text B},l}^{\text{post}}(t)/L,
\end{align}
so that ${\text {Proj}}[q_{\text B}^t({\mathbf z}_l)]$ is approximated as
\begin{align}\label{qtilde}
{\text {Proj}}[q_{\text B}^t({\mathbf z}_l)]\approx {\mathcal N}({\mathbf z}_{l};{\mathbf z}_{{\text B},l}^{\text{post}}(t),{v}_{\text B}^{\text{post}}(t){\mathbf I}_M)\triangleq \tilde{q}_{\text B}^t({\mathbf z}_l).
\end{align}
As a result, the message $m_{{\mathbf z}_l\rightarrow \delta_z}^t({\mathbf z}_l)$ from the variable node ${\mathbf Z}$ to the factor node $\delta({\mathbf Z}-{\mathbf A}({\boldsymbol \theta}_A){\mathbf X})$ can be calculated (componentwise) as
\begin{subequations}\label{delta}
\begin{align}
m_{{\mathbf z}_l\rightarrow \delta_z}^t({\mathbf z}_l)\propto \frac{{\mathcal N}({\mathbf z}_{l};{\mathbf z}_{{\text B},l}^{\text{post}}(t),{v}_{{\text B}}^{\text{post}}(t){\mathbf I}_M)}{{\mathcal N}({\mathbf z}_l;{\mathbf z}_{{\text A},l}^{\text {ext}}(t),{v}_{{\text A}}^{\text {ext}}(t){\mathbf I}_M)}&\propto {\mathcal N}({\mathbf z}_{l};{\mathbf z}_{{\text B},l}^{\text {ext}}(t),{v}_{{\text B}}^{\text {ext}}(t){\mathbf I}_M),
\end{align}
\end{subequations}
where the extrinsic means ${\mathbf z}_{{\text B},l}^{\text {ext}}(t)$ and variances ${v}_{{\text B}}^{\text {ext}}(t)$ are
\begin{align}
&{v}_{{\text B}}^{\text {ext}}(t)=\left(\frac{1}{{v}_{{\text B}}^{\text{post}}(t)}-\frac{1}{{v}_{{\text A}}^{\text {ext}}(t)}\right)^{-1},\label{extB_var}\\
&{\mathbf z}_{{\text B},l}^{\text{ext}}(t)={v}_{{\text B}}^{\text{ext}}(t)\left(\frac{{\mathbf z}_{{\text B},l}^{\text{post}}(t)}{{v}_{{\text B}}^{\text{post}}(t)}-\frac{{\mathbf z}_{{\text A},l}^{\text {ext}}(t)}{{v}_{{\text A}}^{\text {ext}}(t)}\right).\label{extB_mean}
\end{align}
To learn the unknown parameter ${\boldsymbol \theta}_Y$, EM can be adopted \cite{EMAMP}, i.e.,
\begin{align}\label{updatethetaY}
{\boldsymbol \theta}_Y(t)&=\underset{{\boldsymbol \theta}_Y}{\operatorname{argmax}}~{\text E}\left[\log p({\mathbf Y}|{\mathbf Z};{\boldsymbol \theta}_Y){m}_{\delta_z\rightarrow {\mathbf Z}}^{t-1}({\mathbf Z})|{\tilde{q}_{\text B}^t({\mathbf z}_l)}\right]\notag\\
&=\underset{{\boldsymbol \theta}_Y}{\operatorname{argmax}}~\sum\limits_{l=1}^L{\text E}\left[\log p({\mathbf y}_l|{\mathbf z}_l;{\boldsymbol \theta}_Y)|{ \tilde{q}_{\text B}^t({\mathbf z}_l)}\right],
\end{align}
where ${ \tilde{q}_{\text B}^t({\mathbf z}_l)}$ is given by (\ref{qtilde}).
\subsection{BAd-VAMP module}
\begin{figure}
  \centering
  \includegraphics[width=8cm]{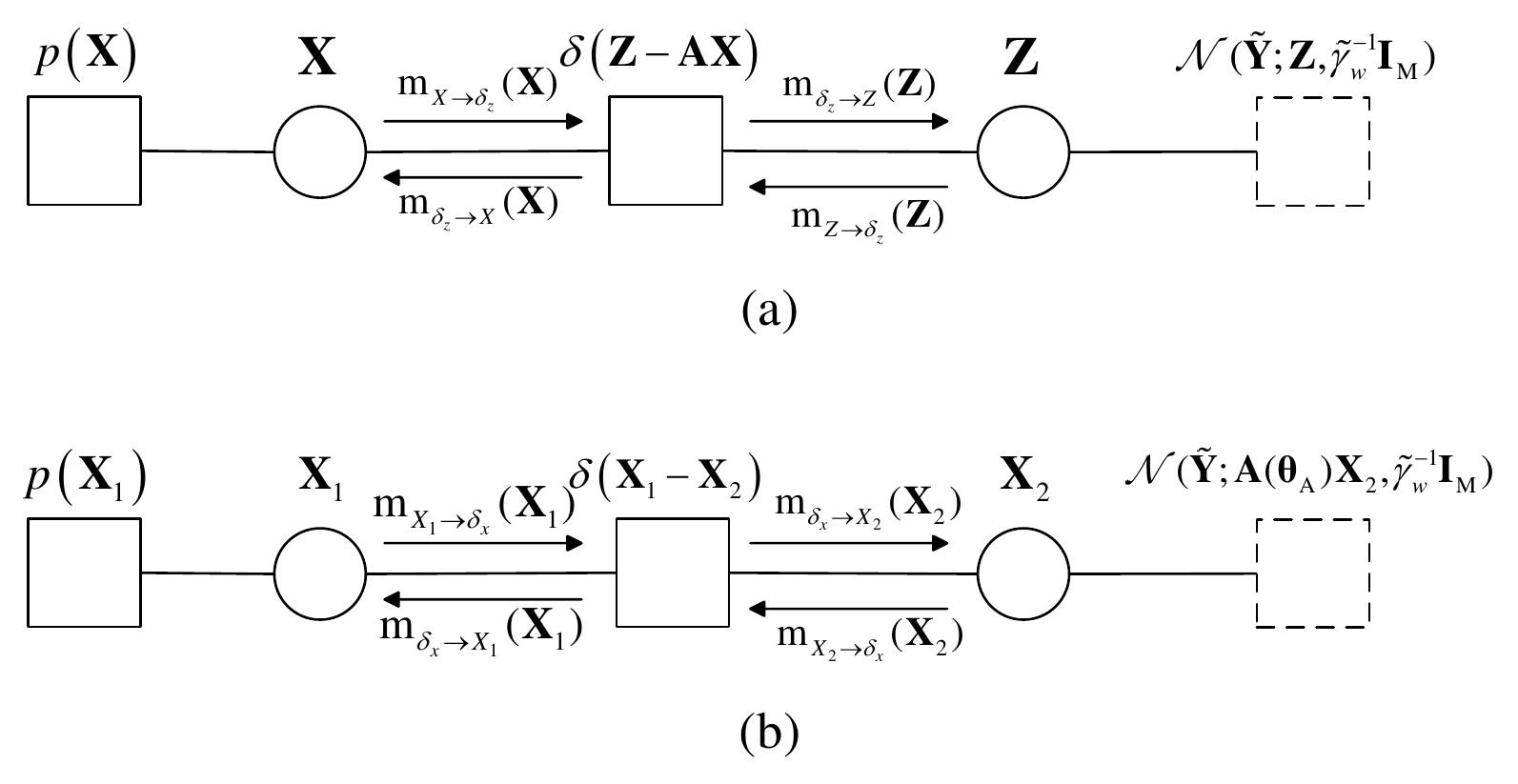}\\
  \caption{Two equivalent factor graphs for the pseudo linear observation model (\ref{pseudolinear}). Note that Fig. \ref{module2} (a) is the proposed factor graph which novelly introduces the delta function, and Fig. \ref{module2} (b) is the factor graph proposed by \cite{VAMP}. }\label{module2}
\end{figure}
As shown in (\ref{delta}), the message $m_{{\mathbf z}_l\rightarrow \delta_z}^t({\mathbf z}_l)$  from the variable node ${\mathbf Z}$ to the factor node $\delta({\mathbf Z}-{\mathbf A}({\boldsymbol \theta}_A){\mathbf X})$ follows Gaussian distribution ${\mathcal N}({\mathbf z}_l;{\mathbf z}_{{\text B},l}^{\text{ext}}(t),{v}_{{\text B}}^{\text{ext}}(t){\mathbf I}_M)$. Referring to the definition of the $\delta(\cdot)$ for the factor node, we obtain a pseudo linear observation equation as
\begin{align}\label{pseudolinear}
\tilde{\mathbf y}_l(t)={\mathbf A}({\boldsymbol \theta}_A){\mathbf x}_l+\tilde{\mathbf w}_l, \quad l = 1,\cdots,L,
\end{align}
where $\tilde{\mathbf y}_l(t)\triangleq {\mathbf z}_{{\text B},l}^{\text{ext}}(t)$, $\tilde{\mathbf w}_l\sim {\mathcal N}(\tilde{\mathbf w}_l;{\mathbf 0};\tilde{\gamma}_w^{-1}(t){\mathbf I}_M)$ and $\tilde{\gamma}_w(t)\triangleq 1/{v}_{{\text B}}^{\text{ext}}(t)$. The factor graph corresponding to (\ref{pseudolinear}) is shown Fig. \ref{module2}, where the dash square is used to indicate pseudo observations. As a result, the BAd-VAMP algorithm \cite{BAdVAMP} for the standard bilinear recovery problem can be applied. For completeness and ease of reference, we present the derivation of BAd-VAMP in \cite{BAdVAMP} based on the factor graph shown in Fig. \ref{module2} (b), in which  replicas of ${\mathbf X}$ are introduced, i.e., ${\mathbf X}_1={\mathbf X}_2={\mathbf X}$. In the following, let ${\mathbf x}_{1,l}$ and ${\mathbf x}_{2,l}$ denote the $l$ th column of ${\mathbf X}_1$ and ${\mathbf X}_2$, respectively. Assume that the message $\{m_{\delta_x\rightarrow {\mathbf x}_{2,l}}^{t-1}({\mathbf x}_{2,l})\}_{l=1}^L$ transmitted from the factor node $\delta({\mathbf X}_1-{\mathbf X}_2)$ to the variable node ${\mathbf X}_2$ is
\begin{align}\label{init2}
m_{\delta_x\rightarrow {\mathbf x}_{2,l}}^{t-1}({\mathbf x}_{2,l})= {\mathcal N}({\mathbf x}_{2,l};{\mathbf r}_{2,l}({t-1});{\gamma}_{2,l}^{-1}(t-1){\mathbf I}_N),
\end{align}
where $\delta_x$ refers to the factor node $\delta({\mathbf X}_1-{\mathbf X}_2)$. Note that $m_{\delta_x\rightarrow {\mathbf x}_{2,l}}^{t-1}({\mathbf x}_{2,l})$ can be viewed as the prior of ${\mathbf x}_{2,l}$. Combining the pseudo observation equation (\ref{pseudolinear}) with ${\boldsymbol \theta}_A(t-1)$, the linear MMSE (LMMSE) estimate of ${\mathbf x}_{2,l}$ is performed and the posterior distribution of ${\mathbf x}_{2,l}$ is obtained as
\begin{align}\label{q2}
q_2^t({\mathbf X}_2)=\prod\limits_{l=1}^L{\mathcal N}({\mathbf x}_{2,l};\hat{\mathbf x}_{2,l}(t),{\boldsymbol \Xi}_{{\mathbf x}_{2,l}}(t))\triangleq\prod\limits_{l=1}^Lq_2^t({\mathbf x}_{2,l}),
\end{align}
where the posterior mean $\hat{\mathbf x}_{2,l}(t)$ and covariance matrix ${\boldsymbol \Xi}_{{\mathbf x}_{2,l}}(t)$ are
\begin{align}
\hat{\mathbf x}_{2,l}(t)&=(\tilde{\gamma}_{w}(t){\mathbf A}^{\text T}({\boldsymbol \theta}_A(t-1)){\mathbf A}({\boldsymbol \theta}_A(t-1))+{\gamma}_{2,l}(t-1){\mathbf I}_N)^{-1}\notag\\
&\times(\tilde{\gamma}_w(t){\mathbf A}^{\text T}({\boldsymbol \theta}_A(t-1))\tilde{\mathbf y}_l(t)+{\gamma}_{2,l}(t-1){\mathbf r}_{2,l}(t-1)),\label{LMMSEa}\\
&{\boldsymbol \Xi}_{{\mathbf x}_{2,l}}(t)=(\tilde{\gamma}_w(t){\mathbf A}^{\text T}({\boldsymbol \theta}_A(t-1)){\mathbf A}({\boldsymbol \theta}_A(t-1))+{\gamma}_{2,l}(t-1){\mathbf I}_N)^{-1}.\label{LMMSEb}
\end{align}
In addition, the EM algorithm is incorporated to learn ${\boldsymbol \theta}_A$ and update the pseudo noise precision $\tilde{\gamma}_w$, i.e.,
\begin{subequations}\label{EM2}
\begin{align}
&{\boldsymbol \theta}_A(t)\notag\\
&=\underset{{\boldsymbol \theta}_A}{\operatorname{argmax}}~{\text E}[\log p_{\tilde{\mathbf Y}(t)|{\mathbf X}_2}(\tilde{\mathbf Y}(t)|{\mathbf X}_2;{\boldsymbol \theta}_A,\tilde{\gamma}_w(t))|q_2^t({\mathbf X}_2)],\label{EM2a}\\
&\tilde{\gamma}_w(t)\notag\\
&=\underset{\tilde{\gamma}_w}{\operatorname{argmax}}~{\text E}[\log p_{\tilde{\mathbf Y}(t)|{\mathbf X}_2}(\tilde{\mathbf Y}(t)|{\mathbf X}_2;{\boldsymbol \theta}_A(t-1),\tilde{\gamma}_w)|q_2^t({\mathbf X}_2)].\label{EM2b}
\end{align}
\end{subequations}
Specifically, for the affine-linear model ${\mathbf A}({\boldsymbol \theta}_A)={\mathbf A}_0+\sum\limits_{i=1}^G{\theta}_{A,i}{\mathbf A}_i$, the detailed expression of estimating ${\boldsymbol \theta}_A$ and $\tilde{\gamma}_w$ are given by \cite{BAdVAMP}
\begin{subequations}
\begin{align}
&{\boldsymbol \theta}_A(t)=({\mathbf H}(t))^{-1}{\boldsymbol \beta}^t,\label{thetaA}\\
&1/\tilde{\gamma}_w(t)=\frac{1}{ML}(\|\tilde{\mathbf Y}(t)-{\mathbf A}({\boldsymbol \theta}_A(t-1))\hat{\mathbf X}_2(t)\|_{\text F}^2+{\text {tr}}\{{\mathbf A}({\boldsymbol \theta}_A(t-1)){\boldsymbol \Xi}_{\mathbf X}(t){\mathbf A}^{\text T}({\boldsymbol \theta}_A(t-1))\}),\label{gammaw}
\end{align}
\end{subequations}
where
\begin{subequations}
\begin{align}
&[{\mathbf H}(t)]_{ij}={\text{tr}}\{{\mathbf A}_j^{\text T}{\mathbf A}_i({\boldsymbol \Xi}_{\mathbf X}(t)+\hat{\mathbf X}_2(t)\hat{\mathbf X}_2^{\text T}(t))\},\\
&[{\boldsymbol \beta}^t]_i={\text{tr}}\{\tilde{\mathbf Y}^{\text T}(t){\mathbf A}_i\hat{\mathbf X}_2(t)\}-{\text{tr}}\{{\mathbf A}_0^{\text T}{\mathbf A}_i({\boldsymbol \Xi}_{\mathbf X}(t)+\hat{\mathbf X}_2(t)\hat{\mathbf X}_2^{\text T}(t))\},
\end{align}
\end{subequations}
${\boldsymbol \Xi}_{\mathbf X}(t)=\sum\limits_{l=1}^L{\boldsymbol \Xi}_{{\mathbf x}_{2,l}}(t)$ and ${\boldsymbol \Xi}_{{\mathbf x}_{2,l}}(t)$ is given by (\ref{LMMSEb}).

The message $m_{{\mathbf x}_{2,l} \rightarrow\delta_x }^t({\mathbf x}_{2,l})$ from the variable node ${\mathbf X}_2$ to the delta node $\delta({\mathbf X}_1-{\mathbf X}_2)$ is calculated as
\begin{align}\label{r1lcal}
m_{{\mathbf x}_{2,l} \rightarrow\delta_x }^t({\mathbf x}_{2,l})\propto \frac{{\text {Proj}}[q_2^t({\mathbf x}_{2,l})]}{m_{\delta_x\rightarrow {\mathbf x}_{2,l}}^{t-1}({\mathbf x}_{2,l})},
\end{align}
where $q_2^t({\mathbf x}_{2,l})$ is defined in (\ref{q2}). Projecting the posterior distribution $q_2^t({\mathbf x}_{2,l})$ to the Gaussian distribution with scalar covariance matrix yields
\begin{align}\label{q2proj}
{\text {Proj}}[q_2^t({\mathbf x}_{2,l})]\propto {\mathcal N}({\mathbf x}_{2,l};\hat{\mathbf x}_{2,l}(t),\eta_{2,l}^{-1}(t)\mathbf I_N),
\end{align}
where
\begin{align}\label{eta2l}
\eta_{2,l}^{-1}(t)={{\text{tr}}({\boldsymbol \Xi}_{{\mathbf x}_{2,l}}(t))}/N.
\end{align}
Substituting (\ref{q2proj}) in (\ref{r1lcal}), we obtain
\begin{align}\label{r1lfinal}
m_{{\mathbf x}_{2,l} \rightarrow\delta_x }^t({\mathbf x}_{2,l})\propto {\mathcal N}({\mathbf x}_{2,l};{\mathbf r}_{1,l}(t),\gamma_{1,l}^{-1}(t)\mathbf I_N),
\end{align}
where
\begin{align}
&\gamma_{1,l}(t)=\eta_{2,l}(t)-\gamma_{2,l}(t-1),\label{gamma1l}\\
&{\mathbf r}_{1,l}(t)=(\eta_{2,l}(t)\hat{\mathbf x}_{2,l}(t)-\gamma_{2,l}(t-1){\mathbf r}_{2,l}(t-1))/\gamma_{1,l}(t).\label{r1l}
\end{align}
According to the definition of the factor node $\delta({\mathbf X}_1-{\mathbf X}_2)$, the message $m_{\delta_x \rightarrow {\mathbf x}_{1,l}}^t({\mathbf x}_{1,l})$ satisfies
\begin{subequations}
\begin{align}
m_{\delta_x \rightarrow {\mathbf x}_{1,l}}^t({\mathbf x}_{1,l})=m_{{\mathbf x}_{2,l} \rightarrow\delta_x }^t({\mathbf x}_{2,l})|_{{\mathbf x}_{2,l}={\mathbf x}_{1,l}}= {\mathcal N}({\mathbf x}_{1,l};{\mathbf r}_{1,l}(t),\gamma_{1,l}^{-1}(t)\mathbf I_N).
\end{align}
\end{subequations}
Combining the prior $p({\mathbf X}_1;{\boldsymbol \theta}_X)$ with ${\boldsymbol \theta}_X(t-1)$, the posterior mean and variances of ${\mathbf X}_1$ are calculated as
\begin{align}
&\eta_{1,l}^{-1}(t)=<{\text {Var}}[{\mathbf x}_{1,l}|q_{1,l}^t({\mathbf x}_{1,l})]>,\label{inputdenoisingvar}\\
&\hat{\mathbf x}_{1,l}(t)={\text E}[{\mathbf x}_{1,l}|q_{1,l}^t({\mathbf x}_{1,l})],\label{inputdenoisingmean}
\end{align}
where
\begin{align}\label{q1lt}
q_{1,l}^t({\mathbf x}_{1,l})\propto p({\mathbf x}_{1,l}){\mathcal N}({\mathbf x}_{1,l};{\mathbf r}_{1,l}(t),\gamma_{1,l}^{-1}(t)\mathbf I_N).
\end{align}
To learn the unknown parameters ${\boldsymbol \theta}_X(t)$ and $\gamma_{1,l}(t)$, EM algorithm is applied in the inner iterations \cite{BAdVAMP}, i.e.,
\begin{align}\label{LearnthetaX}
&{\boldsymbol \theta}_X(t)=\underset{{\boldsymbol \theta}_X}{\operatorname{argmax}}~{\text E}[\log p_{{\mathbf X}}({\mathbf X}_1;{\boldsymbol \theta}_X)|q_1^t({\mathbf X}_1)],
\end{align}
and
\begin{align}
\gamma_{1,l}(t)&=\underset{\gamma_{1,l}}{\operatorname{argmax}}~{\text E}[\log p({\mathbf r}_{1,l}(t)|{\mathbf x}_l;\gamma_{1,l})|q_1^t({\mathbf X}_1)]\\
&=\left\{\frac{1}{N}\|\hat{\mathbf x}_{1,l}({t})-{\mathbf r}_{1,l}({t})\|^2+\frac{1}{\eta_{1,l}(t)}\right\}^{-1}.
\end{align}
Now the message $m_{{\mathbf x}_{1,l}\rightarrow \delta_x}^t({\mathbf x}_{1,l})$ from the variable node ${\mathbf X}_1$ to the factor node $\delta({\mathbf X}_1-{\mathbf X}_2)$ is calculated as
\begin{subequations}\label{r2l}
\begin{align}
m_{{\mathbf x}_{1,l}\rightarrow \delta_x}^t({\mathbf x}_{1,l})&\propto \frac{{\text {Proj}}[q_{1,l}^t({\mathbf x}_{1,l})]}{m_{\delta_x \rightarrow {\mathbf x}_{1,l}}^t({\mathbf x}_{1,l})}\propto {\mathcal N}({\mathbf x}_{1,l};{\mathbf r}_{2,l}(t),\gamma_{2,l}^{-1}(t)\mathbf I_N),
\end{align}
\end{subequations}
where $q_{1,l}^t({\mathbf x}_{1,l})$ is (\ref{q1lt}), ${\mathbf r}_{2,l}(t)$ and $\gamma_{2,l}^{-1}(t)$ are given by
\begin{align}
&\gamma_{2,l}(t)=\eta_{1,l}(t)-\gamma_{1,l}(t),\label{gamma2l}\\
&{\mathbf r}_{2,l}(t)=(\eta_{1,l}(t){\mathbf x}_{1,l}(t)-\gamma_{1,l}(t){\mathbf r}_{1,l}(t))/\gamma_{2,l}(t).\label{r2l}
\end{align}
According to the definition of the factor node $\delta({\mathbf X}_1-{\mathbf X}_2)$, the message $m_{\delta_x\rightarrow{\mathbf x}_{2,l} }^t({\mathbf x}_{2,l})$ from the factor node $\delta({\mathbf X}_1-{\mathbf X}_2)$ to the variable node ${\mathbf X}_2$ is $m_{\delta_x\rightarrow{\mathbf x}_{2,l} }^t({\mathbf x}_{2,l})= {\mathcal N}({\mathbf x}_{2,l};{\mathbf r}_{2,l}(t),\gamma_{2,l}^{-1}\mathbf I_N)$,
which closes the BAd-VAMP algorithm.
\subsection{Messages from BAd-VAMP module to MMSE module}\
After performing BAd-VAMP for one or more iterations, we now focus on how to calculate the extrinsic message $m_{\delta_z\rightarrow {\mathbf z}_l}^{t}({\mathbf z}_l)$ from the BAd-VAMP module to the component-wise MMSE module. Referring to the original factor graph shown in Fig. \ref{module} (a), according to EP, the extrinsic message $m_{\delta_z\rightarrow {\mathbf z}_l}^{t}({\mathbf z}_l)$ can be calculated as
\begin{subequations}\label{mmm}
\begin{align}
m_{\delta_z\rightarrow {\mathbf z}_l}^{t}({\mathbf z}_l)\propto \frac{{\text {Proj}}\left[\int_{{\mathbf x}_l}\delta({\mathbf z}_l-{\mathbf A}{\mathbf x}_l)m_{{\mathbf x}_l\rightarrow \delta_z}^t({\mathbf x}_l){\text d}{\mathbf x}_l
m_{{\mathbf z}_l\rightarrow\delta_z }^{t}({\mathbf z}_l)\right]}{m_{{\mathbf z}_l\rightarrow \delta_z}^{t}({\mathbf z}_l)}\triangleq \frac{{\text {Proj}}[q_{\text A}^{t}({\mathbf z}_l)]}{m_{{\mathbf z}_l\rightarrow \delta_z}^{t}({\mathbf z}_l)}.
\end{align}
\end{subequations}
In BAd-VAMP, as shown in the above subsection B, we have already obtained the message $m_{\delta_x\rightarrow{\mathbf x}_{2,l} }^t({\mathbf x}_{2,l})$ from the factor node $\delta({\mathbf X}_1-{\mathbf X}_2)$ to the variable node ${\mathbf X}_2$. It can be seen from Fig. \ref{module2} that the message $m_{{\mathbf x}_l\rightarrow \delta_z}^t({\mathbf x}_l)$ is the same as $m_{\delta_x\rightarrow{\mathbf x}_{2,l} }^t({\mathbf x}_{2,l})$ so that $m_{{\mathbf x}_l\rightarrow \delta_z}^t({\mathbf x}_l)={\mathcal N}({\mathbf x}_{l};{\mathbf r}_{2,l}(t),\gamma_{2,l}^{-1}(t)\mathbf I_N)$. After some algebra, the posterior distribution $q_{\text A}^{t}({\mathbf z}_l)$ of ${\mathbf z}_l$  can be calculated to be Gaussian, i.e., $q_{\text A}^{t}({\mathbf z}_l) = {\mathcal N}({\mathbf z}_l;{\mathbf z}_{{\text A},l}^{\text{post}}(t),{\boldsymbol \Xi}_{{\mathbf z}_l}(t))$, with the covariance matrix and mean vector being
\begin{align}
&{\boldsymbol \Xi}_{{\mathbf z}_l}(t)={\mathbf A}({\boldsymbol \theta}_A(t))\left(\gamma_{2,l}(t){\mathbf I}_N+\tilde{\gamma}_w(t){\mathbf A}^{\text T}({\boldsymbol \theta}_A(t)){\mathbf A}({\boldsymbol \theta}_A(t))\right)^{-1}{\mathbf A}^{\text T}({\boldsymbol \theta}_A(t)),\\
&{\mathbf z}_{{\text A},l}^{\text{post}}(t)={\mathbf A}({\boldsymbol \theta}_A(t))\left(\gamma_{2,l}(t){\mathbf I}_N+\tilde{\gamma}_w(t){\mathbf A}^{\text T}({\boldsymbol \theta}_A(t)){\mathbf A}({\boldsymbol \theta}_A(t))\right)^{-1}(\gamma_{2,l}(t){\mathbf r}_{2,l}(t)+\tilde{\gamma}_w(t){\mathbf A}^{\text T}({\boldsymbol \theta}_A(t))\tilde{\mathbf y}_l(t)).\label{postzmeanz}
\end{align}
Then, the posterior distribution $q_{\text A}^{t}({\mathbf z}_l)$ of ${\mathbf z}_l$ is further projected to Gaussian distribution with scalar covariance matrix, yielding
\begin{align}
{\text {Proj}}[q_{\text A}^{t}({\mathbf z}_l)]={\mathcal N}({\mathbf z}_l;{\mathbf z}_{{\text A},l}^{\text{post}}(t),{v}_{{\text A},l}^{\text{post}}(t){\mathbf I}_M),
\end{align}
where
\begin{align}\label{postz}
&{v}_{{\text A},l}^{\text{post}}(t)={\text{tr}}({\boldsymbol \Xi}_{{\mathbf z}_l}(t))/M,
\end{align}
Moreover, the posterior variances $\{{v}_{{\text A},l}^{\text{post}}(t)\}_{l=1}^L$ are averaged over the index $l$, which leads to
\begin{align}
&{v}_{{\text A}}^{\text{post}}(t)=\sum\limits_{l=1}^L{v}_{{\text A},l}^{\text{post}}(t)/L,\label{postzvarz}
\end{align}
by which ${\text {Proj}}[q_{\text A}^{t}({\mathbf z}_l)]$ is approximated as ${\text {Proj}}[q_{\text A}^{t}({\mathbf z}_l)]\approx {\mathcal N}({\mathbf z}_l;{\mathbf z}_{{\text A},l}^{\text{post}}(t),{v}_{{\text A}}^{\text{post}}(t){\mathbf I}_M)$.
As a result, the message $m_{\delta_z\rightarrow {\mathbf z}_l}^{t+1}({\mathbf z}_l)$ in (\ref{mmm}) becomes
\begin{align}
m_{\delta\rightarrow {\mathbf z}_l}^{t}({\mathbf z}_l)\propto {\mathcal N}({\mathbf z}_l;{\mathbf z}_{{\text A},l}^{\text{ext}}(t),{v}_{{\text A}}^{\text{ext}}(t){\mathbf I}_M),
\end{align}
where
\begin{align}
&{v}_{{\text A}}^{\text{ext}}(t)=\left(\frac{1}{{v}_{{\text A}}^{\text{post}}(t)}-\frac{1}{{v}_{{\text B}}^{\text{ext}}(t)}\right)^{-1},\label{vaext}\\
&{\mathbf z}_{{\text A},l}^{\text{ext}}(t)={v}_{{\text A}}^{\text{ext}}(t)\left(\frac{{\mathbf z}_{{\text A},l}^{\text{post}}(t)}{{v}_{{\text A}}^{\text{post}}(t)}-\frac{{\mathbf z}_{{\text B},l}^{\text{ext}}(t)}{{v}_{{\text B}}^{\text{ext}}(t)}\right),\label{zaext}
\end{align}
which closes the loop of the whole algorithm.

To sum up, the BAd-GVAMP algorithm can be summarized as Algorithm 1.
 \begin{algorithm}[h]
\caption{Bilinear adaptive generalized VAMP (BAd-GVAMP)}
\begin{algorithmic}[1]
\STATE {\bf Initialization}: ${\mathbf z}_{{\text A},l}^{\text{ext}}(0)$, ${v}_{{\text A}}^{\text{ext}}(0)$, ${\mathbf r}_{2,l}(0)$, ${\gamma_{2,l}}(0)$ ${\boldsymbol \theta}_X(0)$, ${\boldsymbol \theta}_A(0)$ and ${\boldsymbol \theta}_Y(0)$.\
\FOR {$t=1,\cdots,T_{\text {outer}}$ }
\STATE Compute the posterior mean and variance of $\mathbf Z$ as ${\mathbf Z}_{\text B}^{\text{post}}(t)$ (\ref{comb_means}), ${v}_{{\text B}}^{\text{post}}(t)$ (\ref{postzb}).
\STATE Compute the extrinsic mean and variance of $\mathbf z$ as ${v}_{{\text B}}^{\text{ext}}(t)$ (\ref{extB_var}), ${\mathbf z}_{{\text B},l}^{\text{ext}}(t)$ (\ref{extB_mean}), and set $\tilde{\mathbf y}_l(t)\triangleq {\mathbf z}_{{\text B},l}^{\text{ext}}(t)$ and $\tilde{\gamma}_{w}(t)\triangleq 1/{v}_{{\text B}}^{\text{ext}}(t)$ in (\ref{pseudolinear}).
\FOR {$\tau=1,\cdots,T_{{\text {inner}},1}$ }
\STATE Perform the LMMSE estimate of ${\mathbf x}_l$, i.e., the posterior means $\hat{\mathbf x}_{2,l}(t)$ and covariance matrix ${\boldsymbol \Xi}_{{\mathbf x}_{2,l}}(t)$ shown in (\ref{LMMSEa}) and (\ref{LMMSEb}).
\STATE Update ${\boldsymbol \theta}_A(t)$ (\ref{EM2a}) and $\tilde{\gamma}_w(t)$ (\ref{EM2b}).
\ENDFOR
\STATE Calculate $\hat{\mathbf x}_{2,l}(t)$ (\ref{LMMSEa}) and $\eta_{2,l}(t)$ (\ref{eta2l}).
\STATE Calculate ${\mathbf r}_{1,l}(t)$ (\ref{r1l}) and $\gamma_{1,l}(t)$ (\ref{gamma1l}).
\FOR {$\tau=1,\cdots,T_{{\text {inner}},2}$ }
\STATE Perform the input denoising operation to obtain the posterior means $\hat{\mathbf x}_{1,l}(t)$ (\ref{inputdenoisingmean}) and variances $\eta_{1,l}^{-1}(t)$ (\ref{inputdenoisingvar}).
\STATE Update ${\boldsymbol \theta}_X$ (\ref{LearnthetaX}).
\STATE Calculate ${\mathbf r}_{2,l}(t)$ (\ref{r2l}) and $\gamma_{2,l}(t)$ (\ref{gamma2l}).
\ENDFOR
\STATE Calculate the posterior means ${\mathbf z}_{{\text A},l}^{\text{post}}(t)$ (\ref{postzmeanz}) and variance ${v}_{{\text A}}^{\text{post}}(t)$ (\ref{postzvarz}).
\STATE Calculate the extrinsic means ${\mathbf z}_{{\text A},l}^{\text{ext}}(t)$ (\ref{zaext}) and variance ${v}_{{\text A}}^{\text{ext}}(t)$ (\ref{vaext}).
\STATE Update ${\boldsymbol \theta}_Y(t)$ as (\ref{updatethetaY}).
\ENDFOR
\STATE Return $\hat{\mathbf X}$ and $\hat{\boldsymbol \Theta}$.
\end{algorithmic}
\end{algorithm}


\subsection{Relation of BAd-GVAMP to BAd-VAMP}
The obtained BAd-GVAMP algorithm is an extension of BAd-VAMP from linear measurements to nonlinear measurements. Intuitively, as shown in Fig \ref{module} (b), BAd-GVAMP iteratively reduces the original generalized bilinear recovery problem to a sequence of standard bilinear recovery problems. In each iteration of BAd-GVAMP, a pseudo linear measurement model is obtained and one iteration of BAd-VAMP is performed \footnote{It is also possible to perform multiple iterations of the BAd-VAMP in a whole single iteration of BAd-GVAMP.}. Note that the message passing schedule of the BAd-VAMP module within BAd-GVAMP is different from the original BAd-VAMP in \cite{BAdVAMP}: in \cite{BAdVAMP} variable de-noising is performed first and then LMMSE, while in the BAd-VAMP module of the proposed BAd-GVAMP, LMMSE is performed first and then variable de-noising. It is worth noting that in the special case of linear measurements, i.e., when $p({\mathbf Y}|{\mathbf Z})$ is Gaussian, i.e., $p({\mathbf y}_{l}|{\mathbf z}_{l})={\mathcal N}({\mathbf y}_{l};{\mathbf z}_{l},\gamma_w^{-1}{\mathbf I}_M)$, the BAd-GVAMP reduces to BAd-VAMP precisely since in such case the extrinsic means ${v}_{{\text B}}^{\text{ext}}(t) $ and variances ${\mathbf z}_{{\text B},l}^{\text{ext}}(t)$ from the MMSE module always satisfy ${v}_{{\text B}}^{\text{ext}}(t)={\gamma}_w^{-1},\forall t,\quad{\mathbf z}_{{\text B},l}^{\text{ext}}(t)={\mathbf y}_l, \forall t,~l$. Thus BAd-GVAMP is consistent with BAd-VAMP under Gaussian output transform.
\section{Numerical Simulation}
In this section, numerical experiments are conducted to investigate the performance of the proposed BAd-GVAMP algorithm. In particular, quantized measurements are considered. As for the inner iteration of BAd-GVAMP in Algorithm 1, we set $T_{{\text {inner}},1}=1$ and $T_{{\text {inner}},2}=2$, which are the same as \cite{BAdVAMP}. In addition, several strategies are proposed to enhance the robustness.
\begin{itemize}
  \item Damping: We perform damping for variables ${\mathbf r}_{1,l}(t)$, $\gamma_{1,l}(t)$, $\gamma_{2,l}(t)$, ${\mathbf r}_{2,l}(t)$. The damping factor is set as $0.8$.
  \item Clipping precisions: Sometimes the variances ${v}_{{\text A}}^{\text{ext}}(t)$ and ${v}_{{\text B}}^{\text{ext}}(t)$ or
  precisions $\{\gamma_{1,l}(t),\gamma_{2,l}(t)\}_{l=1}^L$ can be either negative or too large, we suggest to clip the precisions and variances to the interval $[{\gamma_{\text {min}}},{\gamma_{\text {max}}}]$. In our simulation, we set ${\gamma_{\text {min}}}=10^{-8}$ and ${\gamma_{\text {max}}}=10^{12}$.
\end{itemize}

As for the quantizer, let $Q(\cdot)$ denote a quantization operation. For the quantizer with bit-depth $N_{\text b}$, uniform quantization is adopted with the thresholds being $\tau_i=Z_{\text {min}}+i\Delta, ~i=1,2,\cdots,2^{N_{\text b}}-1$, where $\Delta=(Z_{\text {max}}-Z_{\text {min}})/2^{N_{\text b}}$, $Z_{\text {min}}$ and $Z_{\text {max}}$ denote the minimum and maximum value of $\mathbf Z$. With such uniform quantization, the measurements $\mathbf Y$ becomes ${\mathbf Y}=Q({\mathbf Z}+{\mathbf W})$ where ${W}_{ij}\sim {\mathcal N}({W}_{ij};0,\gamma_w^{-1})$. In this setting, ${\theta}_Y=\{\gamma_w\}$ and is updated approximately as ${\theta}_Y(t)=\tilde{\gamma}_w(t)$.
\subsection{CS with matrix uncertainty from quantized measurements}
Consider the problem ${\mathbf y}=Q({\mathbf A}({\mathbf b}){\mathbf c}+{\mathbf w})$, where ${\mathbf w}\sim {\mathcal N}({\mathbf w};{\mathbf 0},{\mathbf I}_M/\gamma_w)$. Here ${\mathbf A}({\mathbf b})={\mathbf A}_0+\sum\limits_{i=1}^Gb_i{\mathbf A}_i$, where $\{{\mathbf A}_i\}_{i=0}^G\in{\mathbb R}^{M\times N}$ are known, $\mathbf b$ are the unknown uncertainty parameters. We set $\gamma_w$ according to the SNR (dB) defined as ${\text {SNR}}\triangleq 10\log{{\text E\|{\mathbf A}{\mathbf c}\|^2}}/{{\text E}\|{\mathbf w}\|^2}=40$ dB. The uncertainty parameters $\mathbf b$ are drawn from ${\mathcal N}({\mathbf 0},{\mathbf I}_G)$, and $\mathbf c$ is generated with uniformly random support with $K$ nonzero elements from ${\mathcal N}({\mathbf 0},{\mathbf I}_K)$. We set $G=10$ and $K=10$. For one-bit quantization, the debiased normalized mean square error (dNMSE) $\underset{\xi_c}{\operatorname{min}}~10\log ({\|{\mathbf c}-\xi_c\hat{\mathbf c}\|_2}/{\|{\mathbf c}\|_2})$ and $\underset{\xi_b}{\operatorname{min}}~10\log ({\|{\mathbf b}-\xi_b\hat{\mathbf b}\|_2}/{\|{\mathbf b}\|_2})$ are used to characterize the performance. While for multi-bit quantization settings, the NMSEs $10\log ({\|{\mathbf c}-\hat{\mathbf c}\|_2}/{\|{\mathbf c}\|_2})$ and $10\log ({\|{\mathbf b}-\hat{\mathbf b}\|_2}/{\|{\mathbf b}\|_2})$ are used instead. The elements of ${\mathbf A}_0$ are drawn i.i.d. from ${\mathcal N}(0,20)$. To provide a benchmark performance of the BAd-GVAMP algorithm, we also evaluate the oracle performance by assuming known $\mathbf b$ or $\mathbf c$.
\begin{figure}
  \centering
  \includegraphics[width=8cm]{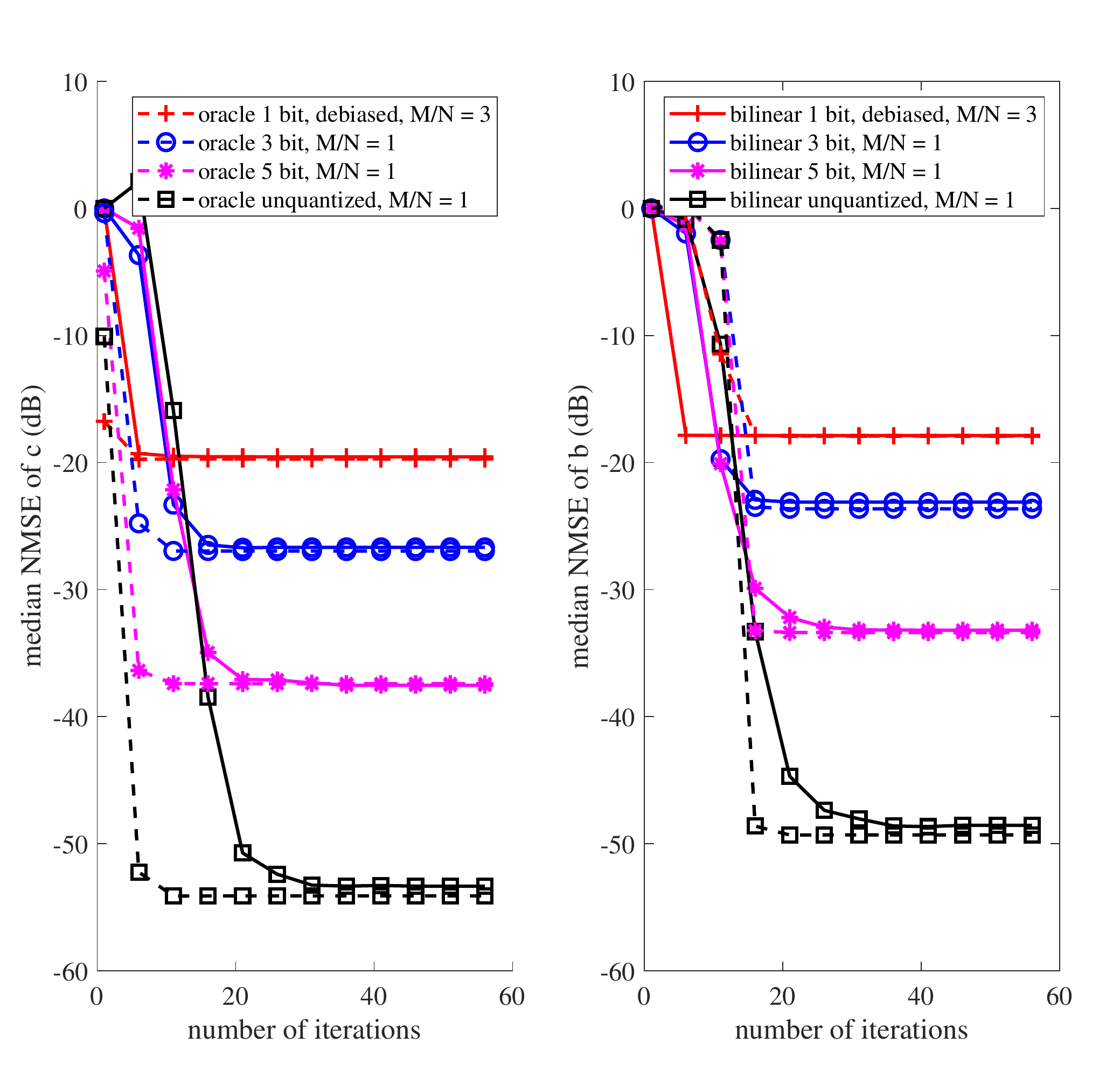}\\
  \caption{Matrix uncertainty scenario: Median NMSE (over 50 Monte Carlo (MC) trials) on signal $\mathbf c$ and uncertainty parameters $\mathbf b$ versus the number of iterations.}\label{nmsevsiter}
\end{figure}

\begin{figure}
  \centering
  \includegraphics[width=8cm]{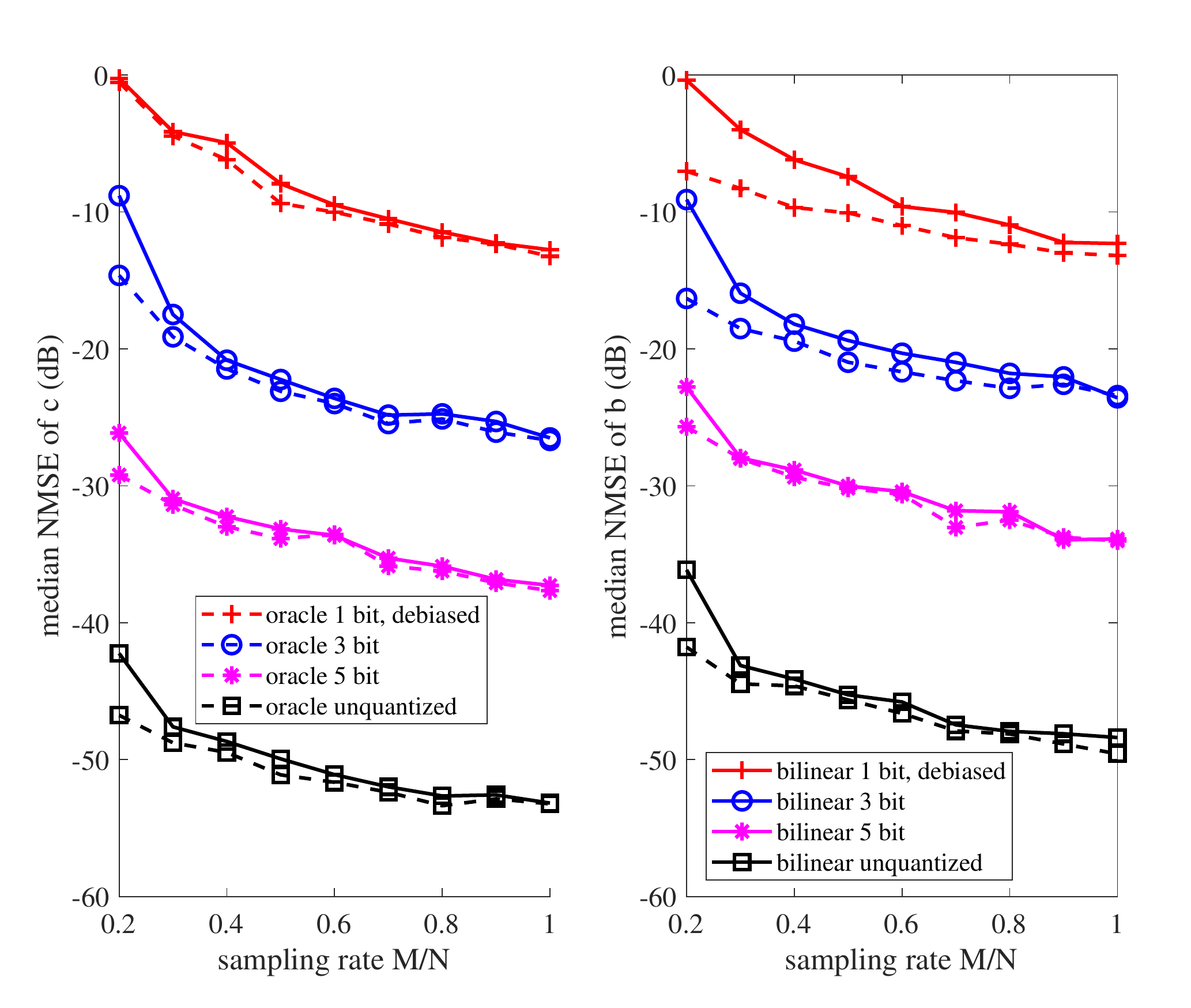}\\
  \caption{Matrix uncertainty scenario: Median NMSE (over 50 MC trials) on signal $\mathbf c$ and uncertainty parameters $\mathbf b$ versus sampling rate $M/N$.}\label{nmseversussamp}
\end{figure}

For the first experiment, we demonstrate the convergence performance of the BAd-GVAMP algorithm. We set the sampling ratio of one-bit quantization as $M/N=3$, while for $3$ bit, $5$ bit and unquantized case, we set $M/N=1$. From Fig. \ref{nmsevsiter}, it can be seen that the BAd-GVAMP algorithm converges after $20\sim 30$ iterations, and its performance is close to the oracle scenario. For the second experiment, the performance versus the sampling rate $M/N$ is investigated. As shown in Fig. \ref{nmseversussamp}, the performance under one-bit quantization is poor for both bilinear and oracle scenarios under low sampling rate. As the sampling rate $M/N$ increases, the performances of BAd-GVAMP algorithm improve. In addition, the proposed algorithm gives near oracle performance for the tested range of $M/N$.
\subsection{Self-Calibration from quantized measurements}
Here we investigate the self-calibration from quantized measurements which aims to recover the $K$-sparse signal vector $\mathbf c$ and the calibration parameters $\mathbf b$ from
\begin{align}
{\mathbf y}&=Q({\text {diag}}({\mathbf H}{\mathbf b}){\boldsymbol \Psi}{\mathbf c}+{\mathbf w})\notag\\
&=Q\left(\left[\sum\limits_{i=1}^Gb_i{\text {diag}}({\mathbf h}_i){\boldsymbol \Psi}\right]{\mathbf c}+{\mathbf w}\right).
\end{align}
with known ${\mathbf H}\in {\mathbb R}^{M\times G}$ and ${\boldsymbol \Psi}\in {\mathbb R}^{M\times N}$. The normalized MSE is defined as
\begin{align}
{\text {NMSE}}\triangleq10\log({\|\hat{\mathbf b}\hat{\mathbf c}^{\text T}-{\mathbf b}{\mathbf c}^{\text T}\|_{\text F}^2}/{\|{\mathbf b}{\mathbf c}^{\text T}\|_{\text F}^2}).\notag
\end{align}
\begin{figure}
  \centering
  \includegraphics[width=8cm]{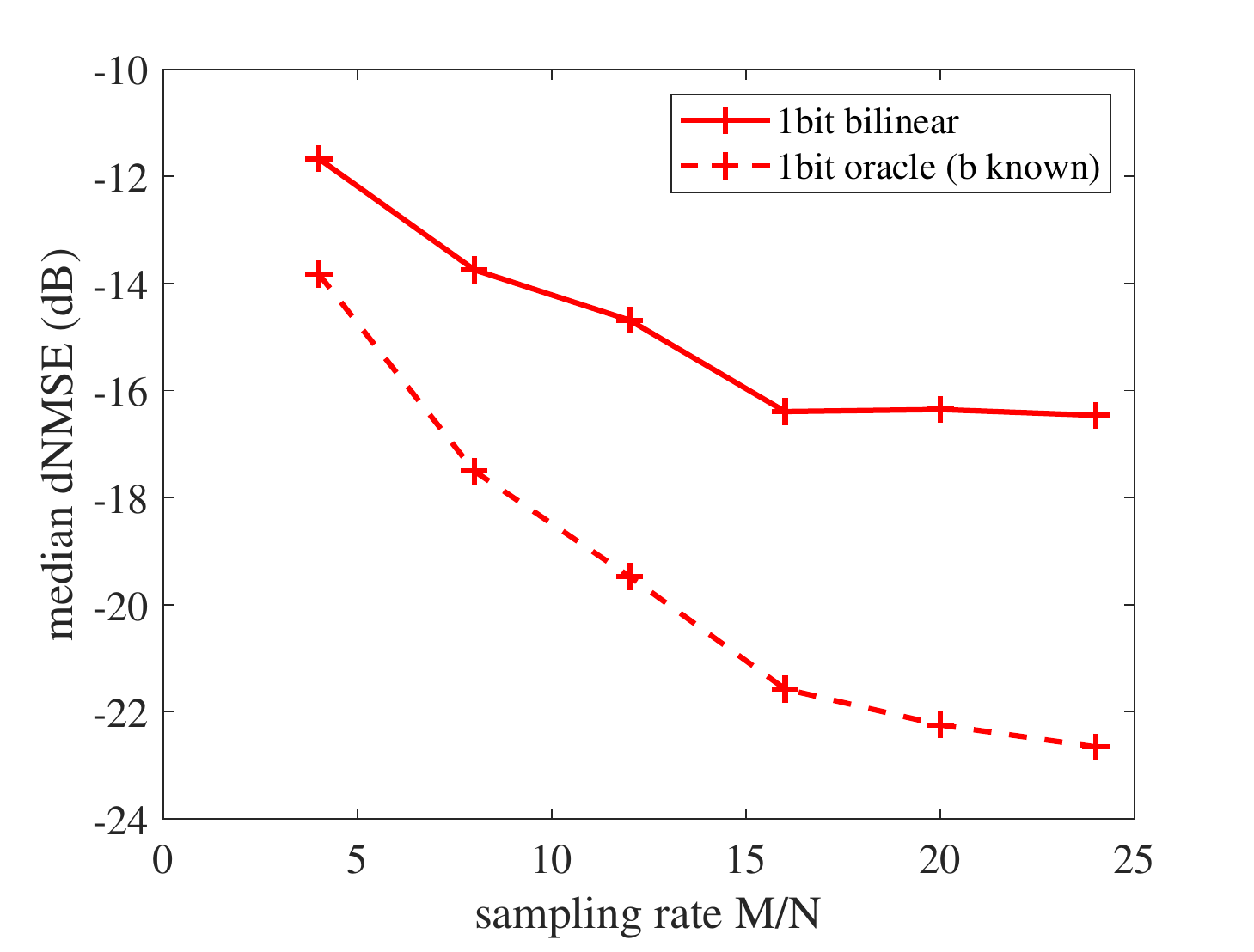}\\
  \caption{Self-Calibration scenario: median NMSE (over 50 MC trials) on signal $\mathbf c$ and uncertainty parameters $\mathbf b$ versus sampling rate $M/N$ under one-bit quantization.}\label{sfnmseversussampb}
\end{figure}

The simulation parameters are set as follows: $K=10$, $G=8$, $M=128$ and ${\text {SNR}}=40$ dB. Here ${\mathbf H}$ is constructed using $Q$ randomly selected columns of the Hadamard matrix, the elements of $\mathbf b$ and ${\boldsymbol \Psi}$ are i.i.d. drawn from ${\mathcal N}(0,1)$, and $\mathbf c$ is generated with $K$ nonzero elements i.i.d. drawn from ${\mathcal N}(0,1)$.
\begin{figure}
  \centering
  \includegraphics[width=8cm]{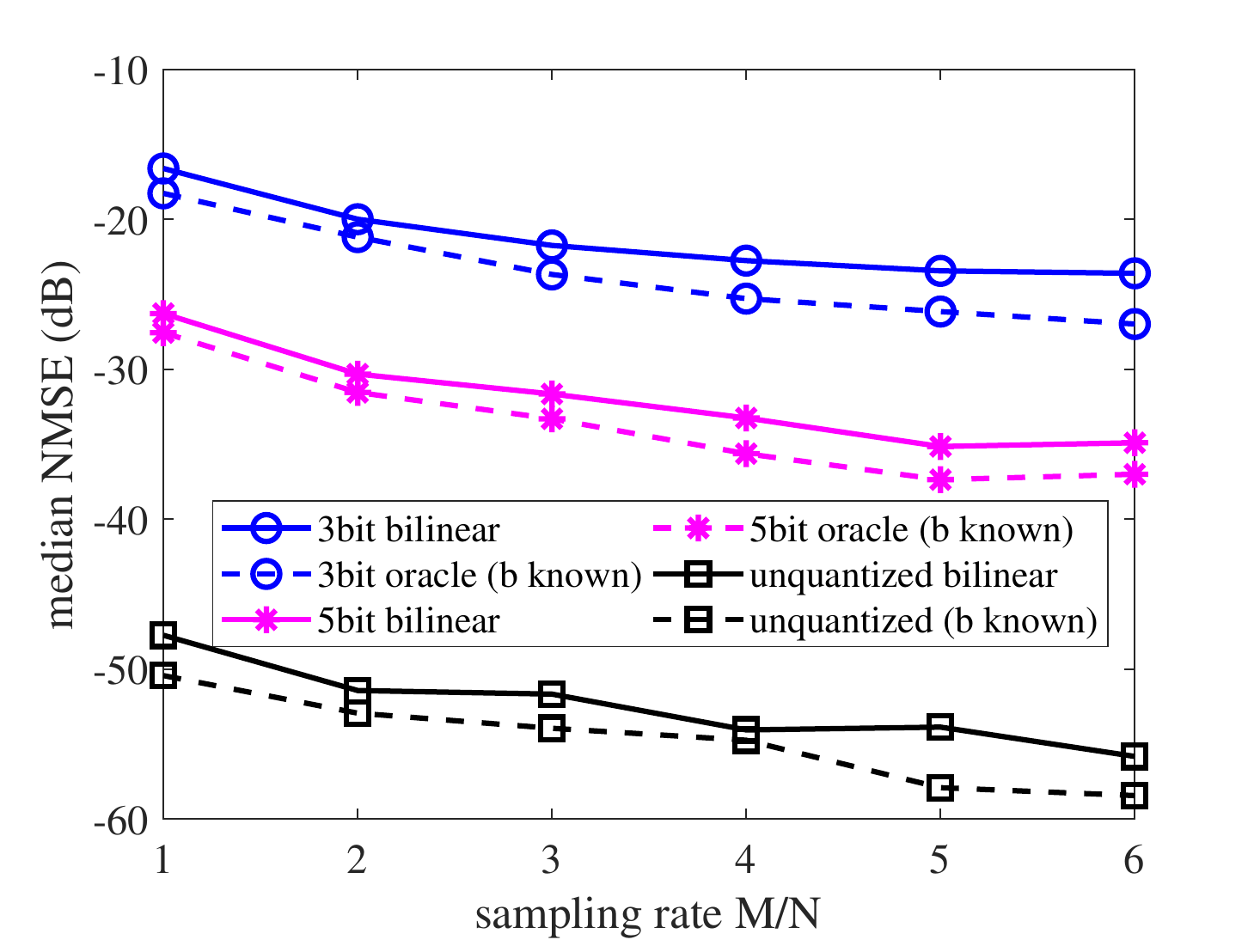}\\
  \caption{Self-Calibration scenario: median NMSE (over 50 MC trials) on signal $\mathbf c$ and uncertainty parameters $\mathbf b$ versus sampling rate $M/N$ under multi-bit quantization.}\label{sfnmseversussampa}
\end{figure}
The NMSE versus the sampling rate $M/N$ are presented in Fig. \ref{sfnmseversussampb} and Fig. \ref{sfnmseversussampa} for one-bit and multi-bit quantization settings, respectively. It can be seen that as the sampling rate increases, the median NMSE decreases. Also, the reconstruction performance improves as the bit-depth increases.
\subsection{Structured dictionary learning from quantized measurements}
The goal of dictionary learning is to find a dictionary matrix ${\mathbf A}\in {\mathbb R}^{M\times N}$ and a sparse matrix ${\mathbf X}\in {\mathbb R}^{N\times L}$ such that ${\mathbf Y}\approx {\mathbf A}{\mathbf X}$ for a given matrix ${\mathbf Y}\in {\mathbb R}^{M\times L}$. We consider structured dictionary $\mathbf A$ such that ${\mathbf A}=\sum\limits_{i=1}^Gb_i{\mathbf A}_i$ with known $\{{\mathbf A}_i\}_{i=1}^G$, where the elements of ${\mathbf A}_i$ and $b_i$ are i.i.d. drawn from ${\mathcal N}(0,1)$ with $G=M=N=64$ in the structured case. Then the measurements are obtained as ${\mathbf Y}=Q({\mathbf A}{\mathbf X}+{\mathbf W})$ such that each column of $\mathbf X$ is $K$ sparse. We set ${\text {SNR}}=40$ dB where SNR is defined as ${\text {SNR}}\triangleq10\log{{\text E}[\|{\mathbf A}{\mathbf X}\|_{\text F}^2]}/{{\text E}[\|{\mathbf W}\|_{\text F}^2]}$. Since the dictionary can not be recovered exactly, the NMSE for the structured case is defined as \cite{BAdVAMP}
\begin{align}
{\text {NMSE}}(\hat{\mathbf A})\triangleq \underset{\lambda\in {\mathbb R}}{\operatorname{min}}~10\log\frac{\|{\mathbf A}-\lambda\hat{\mathbf A}\|_{\text F}^2}{\|{\mathbf A}\|_{\text F}^2}.\notag
\end{align}

\begin{figure}
  \centering
  \includegraphics[width=8cm]{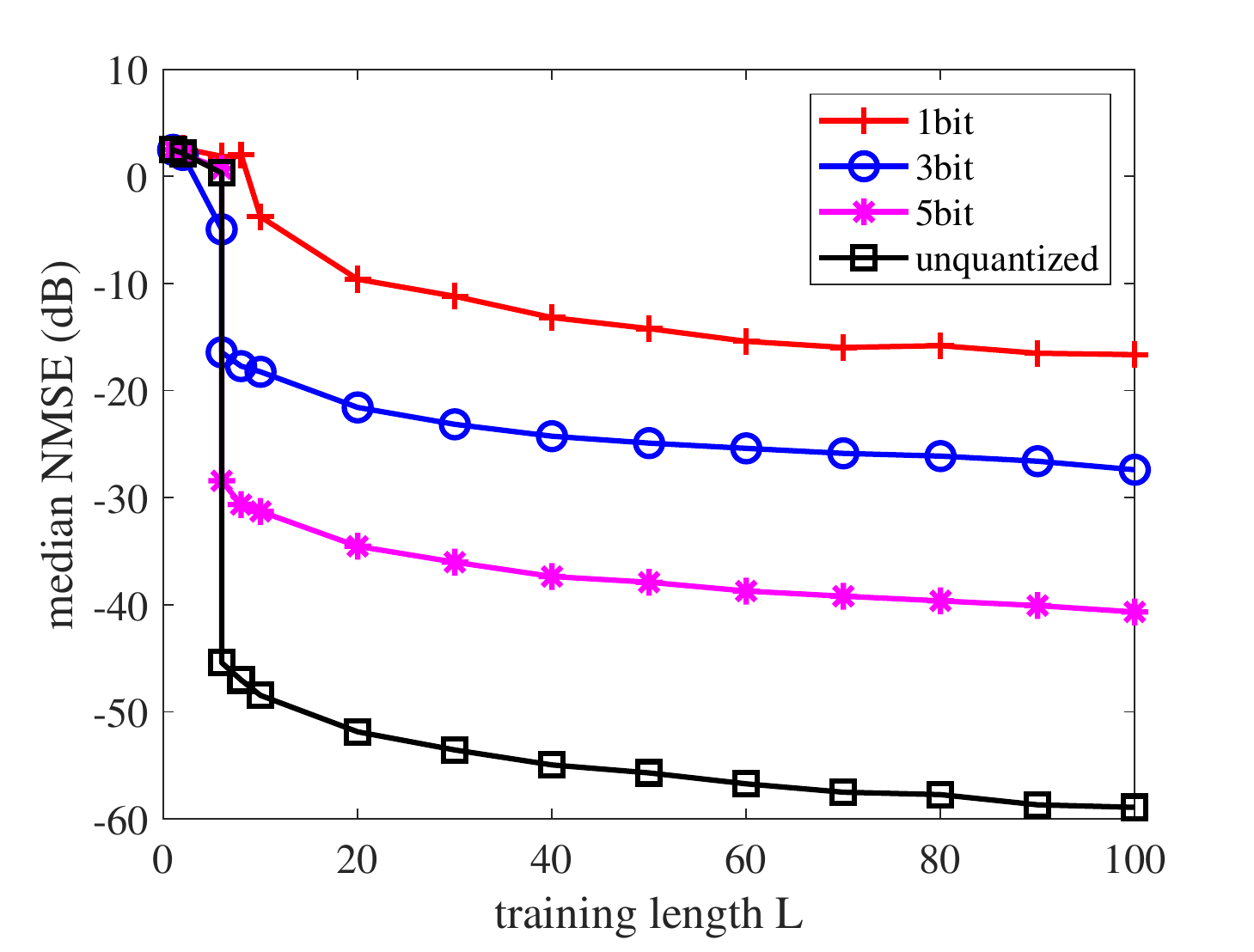}\\
  \caption{Structured dictionary learning scenario: Median NMSE (over 50 MC trials) on dictionary $\mathbf A$ versus the training length $L$.}\label{DLiterfig}
\end{figure}

The median NMSE versus the training length $L$ is shown in Fig. \ref{DLiterfig}. It can be seen that as the training length increases, the NMSE decreases. In addition, the structured dictionary can be learned from one-bit measurements.
\section{Conclusion}
Many problems in science and engineering can be formulated as the generalized bilinear inference problem. To address this problem, this paper proposed a novel algorithm called Bilinear Adaptive Generalized Vector Approximate Message Passing (BAd-GVAMP), which extends the recently proposed BAd-VAMP from linear measurements to nonlinear measurements. In the special case of linear measurements, BAd-GVAMP reduces to the BAd-VAMP. Numerical simulations are conducted for compressed sensing with matrix uncertainty, self-calibration as well as structured dictionary learning from quantized measurements, which demonstrates the effectiveness of the proposed algorithm.

\end{document}